\documentclass[12pt]{article}
\usepackage{graphicx}
\usepackage{amsfonts}
\usepackage{amssymb}
\newcommand{\nin}{\noindent}
\newcommand{\be}{\begin{equation}}
\newcommand{\ee}{\end{equation}}
\newcommand{\bea}{\begin{eqnarray}}
\newcommand{\eea}{\end{eqnarray}}

\newcommand{\nn}{\nonumber\\}

\newcommand{\ol}{\overline}

\begin{document}

\begin{center}

{\Large{\bf{ Schwinger-Dyson approach for a Lifshitz-type Yukawa model}}}

\vspace{0.5cm}

J. Alexandre$^a$\footnote{jean.alexandre@kcl.ac.uk}, K. Farakos$^b$\footnote{kfarakos@central.ntua.gr}, P.
Pasipoularides$^b$\footnote{p.pasip@gmail.com}, A. Tsapalis $^{b,c}$\footnote{a.tsapalis@iasa.gr}

\vspace{0.5cm}

$^a$ Department of Physics, King's College London, WC2R 2LS, UK

$^b$ Department of Physics, National Technical University of
       Athens \\ Zografou Campus, 157 80 Athens, Greece

$^c$ Hellenic Naval Academy, Hatzikyriakou Avenue, Pireaus 185 39, Greece

\vspace{2cm}

{\bf Abstract}

\end{center}

\nin

We consider a 3+1 dimensional field theory at a Lifshitz point for a dynamical critical 
exponent z=3, with a
scalar and a fermion field coupled via a
Yukawa interaction. 
Using the non-perturbative Schwinger-Dyson approach we calculate quantum corrections to the
effective action. We demonstrate that a first order derivative kinetic term as well as a mass
term for the fermion arise dynamically. This signals the restoration of Lorentz symmetry in 
the IR regime of the single fermion model, although for theories with more than one 
fermionic species such a conclusion will require fine-tuning of couplings.
The limitations of the model and our approach are discussed.

\eject

\section{Introduction}

Quantum field theory models, in which the UV behavior is governed by a Lifshitz-type fixed point
have attracted attention recently, as their
renormalization properties appear significantly improved, compared to models with a Lorentz
symmetric Gaussian fixed point.
A novel quantum gravity model, which claims power counting renormalizability, has been formulated
recently by Horava in \cite{Horava:2008ih,
Horava:2009uw}. This scenario is based on an anisotropy between space and time coordinates,
which is expressed via the scalings
$t\rightarrow b^z t$ and  $x\rightarrow b x$, where $z$ is a dynamical critical exponent.
For $z\neq1$ the UV behavior of the model is governed by a nonstandard
Lifshitz fixed point, while for $z=1$ we recover the well known Gaussian fixed point.
Note that in the Horava model, $z=3$ is chosen.

Horava gravity has stimulated an extended research on cosmology and black hole solutions, see for example
\cite{Calcagni:2009ar,Kehagias:2009is,Sotiriou:2009bx,Cai:2009in,Wang:2009yz,Ghodsi:2009zi}. We would like to note that
Horava gravity
is a non-relativistic theory, however it is expected that general relativity
is recovered in the IR limit. Moreover, some possible inconsistencies on Horava gravity have
been remarked in \cite{Charmousis:2009tc,Li:2009bg, Blas:2009qj}, but they
will not be discussed here.

Independently of general relativity,
quantum field theory models in flat space-time with anisotropy have been studied as well.
For example, a thorough study on renormalization properties of models with a Lifshitz-type fixed point,
is presented in \cite{Anselmi:2007ri,Anselmi:2008ry,Anselmi:2008bq,Anselmi:2008bs},
and the Standard Model in this Lorentz violating approach is examined in
\cite{Anselmi:2008bt}. Also, the renormalizability of scalar field theory at the Lifshitz point
is examined in \cite{Visser:2009fg}, and in
\cite{Chen:2009ka} renormalizable models with a Lifshitz fixed point are constructed, whereas a
renormalizable asymptotically free Yang Mills theory,
in 4+1 dimensions, is given in \cite{Horava:2008jf}. As far as dynamical mass generation is concerned,
a four-fermion interaction has been studied in
the framework of Lifshitz-like theories \cite{dhar}, where the authors find a gap equation for the
fermion mass, and the $CP^{N-1}$ model at the
Lifshitz point is discussed in \cite{Das:2009ba}.
In addition, \cite{orlando} shows some perturbative properties of Lifshitz-like theories
containing scalars and fermions,
where an extension of supersymmetry to a Lorentz non-invariant theory is studied.
For a presentation of renormalization group equations in the
case of a scalar field, see \cite{Iengo:2009ix}. Finally, in \cite{Das:2009fb}
a U(1) Gauge theory in 2+1 dimensions with $z=2$ is considered.

The literature mainly deals with perturbative studies, and
our aim here is to study a simple field theory model,
in the framework of the non-perturbative Schwinger-Dyson approach.
In particular, we consider a Lifshitz-type model, in flat space time and in 3+1 dimensions, for a dynamical
critical exponent z=3, with a scalar and a fermion field interacting via a Yukawa coupling.
For the construction of the bare action of the model, we
use only the quadratic marginal operators (kinetic terms), with dimension six,
plus a Yukawa interaction
term with a dimensionful coupling. Note that the construction of more complicated
models, including other marginal and relevant operators (for $z=3$) is possible.
However, in this work we will restrict our study to a Yukawa interaction only,
in order to deal with a tractable system of equations, describing the dynamical generation of mass and Lorentz symmetry
for fermions.

We stress that, for the specific model we examine, the generation of a term of the form  $\lambda i \bar{\psi}\gamma^{\mu}\partial_{\mu}\psi$ is enough for the restoration of Lorentz symmetry in the IR limit of the fermionic sector. However, 
as Ref.\cite{Iengo:2009ix} points out,  the issue of Lorentz symmetry restoration becomes problematic when one considers 
models with more than one species of particles, and we will discuss this point before the conclusion. 

To summarize, this study is based on the Schwinger-Dyson approach, for which
we derive in Appendix A the corresponding equation for the fermion self energy.
The latter is parametrized by two dressed parameters, $m_f$ and $\lambda$, via the operators $m_f^3 \bar{\psi}\psi$ and
$\lambda i \bar{\psi}\gamma^{\mu}\partial_{\mu}\psi$, and the
corresponding self consistent equations are solved.  Note that the parameter $\lambda$
controls the restoration of Lorentz symmetry in the fermionic IR sector.
The coupled evolutions of these two parameters with the Yukawa coupling
is presented in fig.\ref{2}, where it is
found that there exists a critical value for the coupling, above which quantum corrections can generate
simultaneously a Lorentz invariant kinetic term
and a mass for fermions. We also comment on the physical relevance of our model and the
limits of our approximations in Appendix B and Appendix C.

\section{Free systems}

We construct in this section the free scalar and fermion models, and derive the corresponding 
propagators
which will be used for the loop calculations in the next section.

\subsection{Scalar field}

Here we remind the reader the construction of an anisotropic scalar model, in $D+1$ dimensions,
starting with the action
\be\label{scalaraction}
S_b=\frac{1}{2}\int dt d^Dx \left( \dot\phi^2-\phi \left(-\Delta\right)^{z}\phi \right),
\ee
where a dot over a letter represents a time derivative.
The action (\ref{scalaraction}) describes a free scalar theory, with the following mass dimensions
\be
[x^k]=-1~~~~~~~~[t]=-z~~~~~~~~[\phi]=\frac{D-z}{2},
\ee
and leads to the following equation of motion
\be
\ddot\phi+\left(-\Delta\right)^z\phi=0.
\ee
We look for a solution by assuming the separation of variable
\be\label{solb}
\phi(t, \textbf{x})=\xi(t)\exp\{i\textbf{ p}\cdot  \textbf{x}\},
\ee
which leads to
\be
\ddot\xi+(\textbf{p}^{2})^{ z}~\xi=0.
\ee
We
obtain
\be
\xi=\xi_0\exp\left(\pm it \omega \right), \quad \omega=(\textbf{p}^2)^\frac{z}{2},
\ee
where $\xi_0$ is a constant, such that the solutions
of the form of eq.(\ref{solb}) represent plane waves in D+1 dimensions,
and the Feynman propagator for the scalar field, which will be used in order
to calculate loop diagrams, is
\be\label{Gb}
G_b(\omega, \textbf{p})=\frac{i}{\omega^2-(\textbf{p}^2)^z+i\varepsilon},
\ee
where $[\omega]=z$. If we
include a mass term $-\frac{1}{2}m_b^{2z} \phi^2$ in the action of eq.(\ref{scalaraction})
the scalar field propagator is modified as
\be\label{Gbm}
\tilde G_b(\omega, \textbf{p})=\frac{i}{\omega^2-(\textbf{p}^2)^z-m_b^{2z}+i\varepsilon},
\ee
where $[m_b]=1$.

\subsection{Fermionic field}

The action for the free fermionic model is
\be \label{actf}
S_f=\int dt d^Dx\left\lbrace \bar{\psi} i\gamma^0\dot\psi+
\bar{\psi}\left(-\Delta\right)^{\frac{z-1}{2}}\left(i\gamma^k\partial_k\right)\psi\right\rbrace,
\ee
where we have included only quadratic marginal
operators which correspond to a Lifshitz fixed point at the ultraviolet. A dimensional analysis gives
\be
[x^k]=-1~~~~~~~~[t]=-z~~~~~~~~[\psi]=\frac{D}{2},
\ee
and the equation of motion is:
\be
i\gamma^0\dot\psi+\left(-\Delta\right)^{\frac{z-1}{2}}\left(i\gamma^k\partial_k\right)\psi=0.
\ee
We make the following ansatz for the solution of the
above equation
\be
\psi(t, \textbf{x})=\theta(t)\hat\psi_{\textbf{p}}\exp\{i \textbf{p}\cdot \textbf{x}\}.
\label{solf}
\ee
where the
spinor part  $\hat\psi_{\textbf{p}}$ is normalized according to the equation $\hat\psi_{\textbf{p}}^{\dag}\hat\psi_{\textbf{p}}=1$. If we multiply
with the Hermitian conjugate we obtain
\be
\ddot\psi+\left(-\Delta\right)^z\psi=0
\label{con}
\ee
The solution (\ref{solf}) should satisfy eq.(\ref{con}), hence we obtain
\be
\theta(t)=\theta_{0} \exp \left(\pm i t\omega \right) ,\quad \omega=({\textbf{p}}^2)^{\frac{z}{2}}
\ee
where
$\theta_0$ is a constant, such that the solutions (\ref{solf})
represent plane waves in D+1 dimensions. The Feynman propagator for
the fermion field is
\bea \label{Gf} G_f(\omega, \textbf{p})&=&
\frac{i}{\omega \gamma^0-\left(\textbf{p}^2\right)^{\frac{z-1}{2}}(\textbf{p} \cdot\gamma)+i \varepsilon}  \\
&=&i~\frac{\omega \gamma^0-\left(\textbf{p}^2\right)^{\frac{z-1}{2}}(\textbf{p} \cdot \gamma)}
{\omega^2-(\textbf{p}^2)^z+i\varepsilon} \nonumber
\eea
where $[\omega]=z$. We can include the mass term
$-m_f^z \bar{\psi}\psi$ in the action (\ref{actf}),
where $[m_f]=1$, as well as an additional quadratic term\footnote{This term is quadratic in the fermion field, but it is not marginal for $z\ne 1$.} 
$\lambda\bar{\psi}\left(i\gamma^k\partial_k\right)\psi$, where $[\lambda]=z-1$,
such that the fermion propagator is finally
\be \label{Gfmlambda}
\tilde G_f(\omega, \textbf{p})=i~\frac{\omega \gamma^0-\left[\left(\textbf{p}^2\right)^{\frac{z-1}{2}}+\lambda\right] (\textbf{p} \cdot\gamma)+m_f^z}{\omega^2-\left[(\textbf{p}^2)^\frac{z-1}{2}+\lambda\right]^2
\textbf{p}^2 -m_f^{2z}+i \varepsilon}
\ee

\section{Dynamics}

\subsection{Model and Schwinger Dyson equations}

We now consider the simplest interaction between scalars and fermions in the Lifshitz context,
through a Yukawa coupling, and start with the following
bare action
\bea\label{model}
S&=& \int dt d^Dx\Bigg\lbrace \bar{\psi} i\gamma^0\dot\psi+
\bar{\psi}\left(-\Delta\right)^{\frac{z-1}{2}}\left(i\gamma^k\partial_k\right)\psi\\
&&~~~~~~~~~~~~~~~+\frac{1}{2}\dot\phi^2+\frac{1}{2}\phi\left(-\Delta\right)^{z}\phi -\frac{1}{2}m_0^{2z}\phi^2-g\phi\bar{\psi}\psi\Bigg\rbrace,
\nonumber
\eea
where the coupling constant has dimension $[g]=\frac{3z-D}{2}$. In the framework of the gradient
expansion, we will consider quantum
corrections up to the first order in momentum only, such that we will look at the corrections to
the scalar mass, and will allow the dynamical
generation of a fermion mass term $-m_f^3\ol\psi\psi$ and of the additional first order fermionic kinetic term
$\lambda\bar\psi(i\gamma^k\partial_k)\psi$, in order to study the restoration of Lorentz invariance for fermions.

In the action (\ref{model}), we start with a bare scalar mass in order to absorb the only UV
divergence which will appear, as we will see, in the
corrections to the scalar mass. No UV divergence will appear in the fermion self energy, due
to the higher order derivatives, and for this reason
$m_f$ and $\lambda$ can be taken equal to zero in the bare action. We note here that, also
because of higher derivatives, the UV divergence we will
find in the corrections to the scalar mass is logarithmic for $D=z=3$, and not quadratic as
it is in a Lorentz-invariant theory.

\begin{figure}
\begin{center}
\includegraphics[width=13cm]{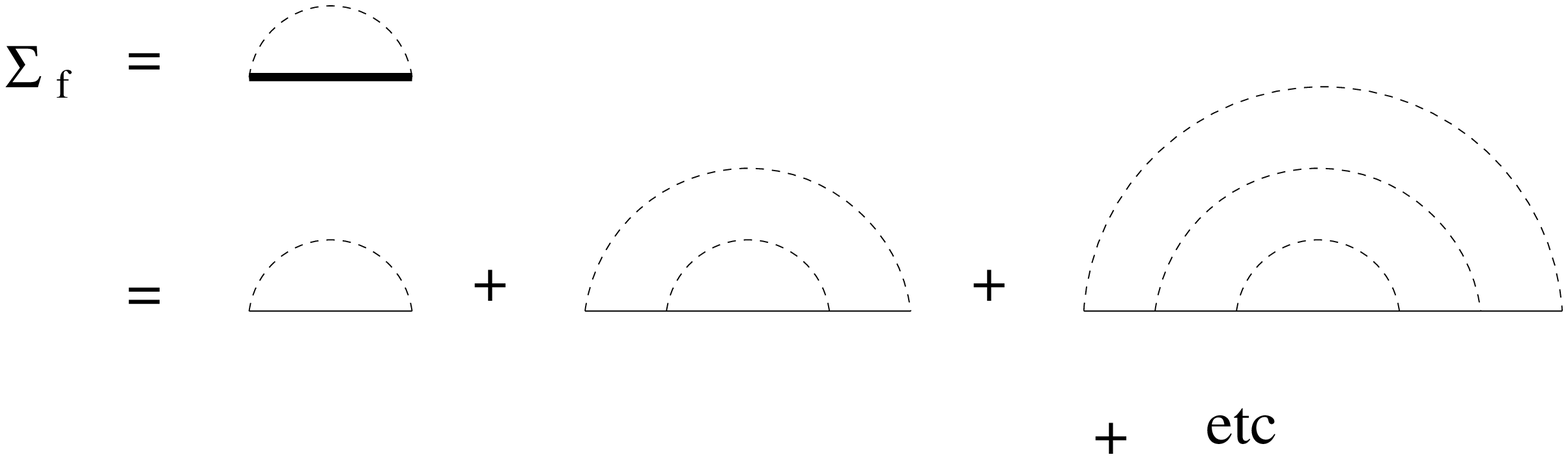}
\end{center}
\caption{\small The fermion self energy given by the Schwinger Dyson equation in the rainbow approximation.
A solid thick line represents the dressed
fermion propagator, a solid thin line the bare fermion propagator, and a dashed line represents the dressed
scalar propagator (which, in our
approximation, is like the bare propagator, but with the renormalized mass instead of the bare one).
As one can see, the fermion self energy is
obtained as a resummation of an infinite number of graphs, which is at the origin of the non-perturbative
feature of the results.}
\label{rainbow}
\end{figure}

We will use here the Schwinger Dyson approach to calculate the fermion and scalar self energies,
which is non-perturbative and represents a resummation
of graphs, avoiding IR divergences, because of the presence of a fermion mass and first order derivative
kinetic term, both generated dynamically.
Also, studies of dynamical mass generation usually lead to a mass which is non-analytical in the coupling
constant, which cannot be found with a naive
loop-expansion, and one therefore needs a non-perturbative approach. We show in the Appendix that the
corresponding Schwinger-Dyson equation for the
fermion self energy $\Sigma_f={\cal G}_f^{-1}-G_f^{-1}$ is
\be\label{SDf}
\Sigma_f=ig{\cal G}_f\Theta {\cal G}_b,
\ee
where ${\cal G}_f,{\cal G}_b$
and $\Theta$ are respectively the dressed fermion propagator, the dressed boson propagator
and the dressed vertex. The equation (\ref{SDf}) is self
consistent, since it displays the dressed quantities on both sides, and therefore corresponds
to a resummation of all quantum corrections (see
fig.(\ref{rainbow})).

Using the exact equation (\ref{SDf}), we can study the dynamical generation of mass and first order
derivative terms for fermions, and we will make the following assumptions:
\begin{itemize}

\item We neglect quantum corrections to the vertex, which corresponds to the so-called ladder or
rainbow approximation \cite{ladder}, and we therefore consider $\Theta\simeq g$.
The corresponding partial resummation provided by the Schwinger-Dyson equations
(\ref{SDf}) is the dominant one for the study of dynamical mass generation\footnote{ There is in principle a infinite tower of Schwinger-Dyson equations, which
are self consistent equations for every $n$-point function, each involving the $n+1$-point function.
A given truncation of this tower of
coupled equations consists then is a specific resummation of graphs for each correlation function.}. We 
show in Appendix B that this approximation is well controlled in the regime where we observe dynamical mass
generation;

\item We also neglect the renormalization of the bare fermion kinetic term, which is consistent in
the framework of the gradient expansion, if we take into account first order derivative corrections
to the fermion dynamics only;

\item Also because of the gradient expansion, we consider a momentum-independent dynamical mass,
since the latter would be quadratic in the momentum.
In addition, the dominant contribution of the
loop integral appearing in the Schwinger-Dyson equation (\ref{SDf}) arises from low momentum,
since no UV divergence occurs in the calculation of the fermion self energy. 
This approximation is discussed in Appendix C.
\end{itemize}
In what follows, we will concentrate on the case $D=z=3$.

\subsection{Scalar sector}

It can be shown, as done in the Appendix for the fermion self energy, that the Schwinger Dyson equation for
the scalar self energy reads
\be
\label{SDb} \Sigma_b=\mbox{Tr}\{{\cal G}_b^{-1}-G_b^{-1}\}=ig\mbox{Tr}\{{\cal G}_f\Theta{\cal G}_f\}.
\ee
As we will see in the next subsection, the
operators $\ol\psi\psi$ and $\bar\psi(i\gamma^k\partial_k)\psi$ will be generated dynamically,
such that we assume here that the dressed fermion
propagator has the form (\ref{Gfmlambda}), ${\cal G}_f=\tilde G_f$, where $m_f$ and $\lambda$
are generated dynamically. The scalar mass, after a Wick
rotation, is then obtained from eq.(\ref{SDb}) for vanishing momentum, which reads
\be \label{mbgapeq}
m_b^6-m_0^6=4g^2\int_{-\infty}^\infty
\frac{d\omega}{2\pi}\int \frac{d^3{\bf p}}{(2\pi)^3}
\frac{\omega^2+({\bf p}^2+\lambda)^2{\bf p}^2-m_f^6}{\left[\omega^2+({\bf p}^2+\lambda)^2
{\bf p}^2+m_f^6\right]^2},
\ee
The integration over $\omega$ leads to a logarithmically-divergent integral over ${\bf p}$:
\bea\label{mb}
m_b^6&=&m_0^6+\frac{g^2}{\pi^2}\int_0^\Lambda \frac{p^4(p^2+\lambda)^2~dp}{[p^2(p^2+\lambda)^2+m_f^6]^{3/2}}\nn &=&m_0^6+\frac{g^2}{\pi^2}\left(\ln\left(\frac{\Lambda}{m_f}\right) +\frac{2\ln2-1}{3}\right)+{\cal O}(\Lambda^{-2}),
\eea
where $\Lambda$ is the cut off in the 3-dimensional ${\bf p}$ space.
Although eq.(\ref{mb}) apparently contains an
IR divergence for $m_f=0$, this divergence is actually not present in this case, if $\lambda\ne0$,
since we have then
\bea
m_b^6&=&m_0^6+\frac{g^2}{2\pi^2}\ln\left(1+\frac{\Lambda^2}{\lambda}\right),
\eea
and $\lambda$ plays the role of IR cut off.

In what follows, the bare mass $m_0$ will be chosen such that the renormalized mass $m_b$ is finite and fixed. This renormalized mass will play the
role of IR cut off for the calculation of the fermion self energy.

\subsection{Fermion sector and self-consistent equations}

The fermion self energy is calculated from the bare propagator (\ref{Gf}) and the dressed propagator
which is assumed to have the form
(\ref{Gfmlambda}), such that
\be\label{sigma}
\Sigma_f({\bf k})=-\lambda({\bf k}\cdot\gamma)-m_f^3.
\ee
Furthermore, if we assume that the dressed
scalar propagator has the form (\ref{Gbm}), ${\cal G}_b=\tilde G_b$, where $m_b$ is
the renormalized, finite scalar mass (\ref{mb}), the right-hand
side of the Schwinger-Dyson equation (\ref{SDf}) is (for vanishing frequency and after a Wick rotation)
\bea\label{rhside}
\Sigma_f({\bf k})&=&-g^2\int_{-\infty}^\infty \frac{d\omega}{2\pi}\int\frac{d^3{\bf p}}{(2\pi)^3}
\frac{i \omega\gamma^0-({\bf p}^2+\lambda)
\left({\bf p}\cdot\gamma\right)+m_f^3}{\omega^2+({\bf p}^2+\lambda)^2{\bf p}^2+m_f^6}\nn
&&~~~~~~~~~~ \times \frac{1}{\omega^2+({\bf p}-{\bf k})^6+m_b^6}.
\eea
This is a convergent integral, and, together with the self energy (\ref{sigma}),
leads to the self consistent equations which must be satisfied by
$\lambda$ and $m_f$: \\
{\it (i)} The equation the fermion dynamical mass should satisfy is obtained by taking the
trace of the Schwinger-Dyson equation (\ref{SDf}), for ${\bf k}=0$:
\be \label{sdm}
m_f^3=\frac{g^2}{(2\pi)^4}\int_{-\infty}^\infty d\omega\int \frac{m_f^3~d^3{\bf p}}{[\omega^2+{\bf p}^6+m_b^6]
[\omega^2+({\bf p}^2+\lambda)^2{\bf p}^2+m_f^6]}.
\ee
If $m_f\ne 0$, the integration over $\omega$ shows that the dynamical mass must satisfy
\be\label{mfgapeq}
\frac{4\pi^2}{g^2}=\int_0^\infty\frac{p^2dp}{A_bA_f(A_b+A_f)},
\ee
where
\bea\label{AbAf} A_b&=&\sqrt{p^6+m_b^6}\nn
A_f&=&\sqrt{p^2(p^2+\lambda)^2+m_f^6}.
\eea
{\it (ii)} The equation for the coefficient $\lambda$ is obtained by expanding the self energy
(\ref{rhside}) in ${\bf k}$, and keeping the linear contribution only in order to identify
it with the corresponding term in eq.(\ref{sigma}). Using the
following equality, valid for any function $f$,
\be
\int d^D{\bf p}({\bf k}\cdot{\bf p})({\bf p}\cdot\gamma)f({\bf p}^2) =\frac{\Omega_D}{D}
({\bf k}\cdot\gamma)\int_0^\infty dp~p^{D+1}f(p^2),
\ee
where $\Omega_D$ is the solid angle in dimension $D$, and identifying the coefficients of
$({\bf k}\cdot\gamma)$ in the Schwinger Dyson equation, we obtain the following self consistent equation
for $\lambda$
\bea\label{sdlambda}
\lambda&=&\frac{g^2}{2\pi^3}\int_{-\infty}^\infty d\omega \int_0^\infty \frac{p^8(p^2+\lambda)dp}{[\omega^2+A_b^2]^2[\omega^2+A_f^2]}\nn
&=&\frac{g^2}{4\pi^2}\int_0^\infty dp~p^8(p^2+\lambda)\frac{2A_b+A_f}{A_b^3A_f(A_b+A_f)^2},
\eea
where $A_f,A_b$ are given in eq.(\ref{AbAf}).
Finally, we are left with the two self-consistent coupled equations (\ref{mfgapeq},\ref{sdlambda}),
which have to be solved simultaneously to find the
parameters $(m_f,\lambda)$ which can be generated dynamically.

\subsection{Numerical analysis and discussion}

In this section, we present our numerical analysis, we comment on the physical relevance of our model and the
limits of our approximation.

\begin{figure}[ht]
\begin{center}
\includegraphics[width=0.8 \textwidth, angle=0]{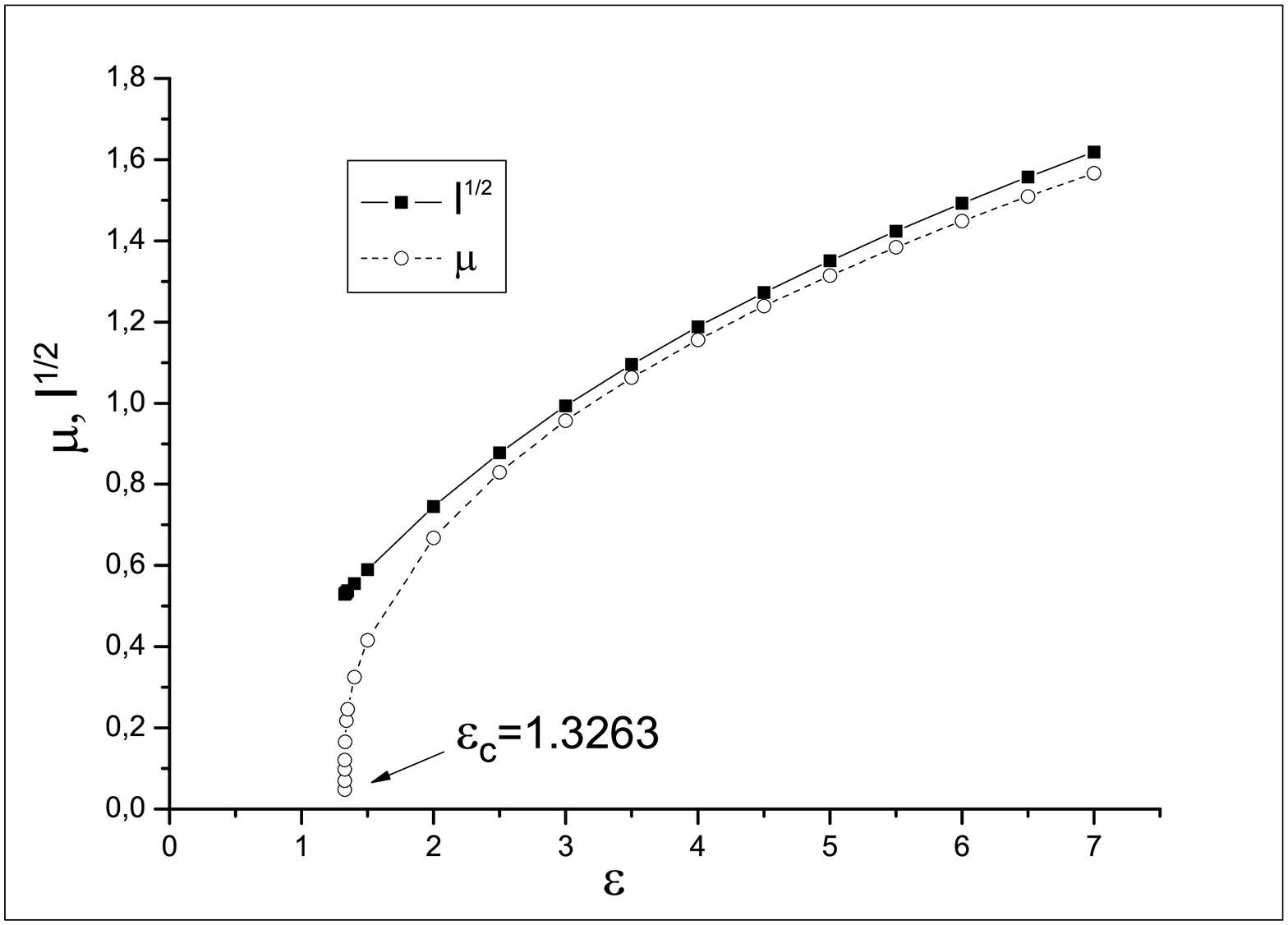}
\end{center}
\caption{\small The parameters $\mu=m_{f}/m_{b}$, $l^{1/2}=\lambda^{1/2}/m_b$ as a function
of $\varepsilon=g/(2\pi m_b^3)$. The system of equations
(\ref{set}) has a unique solution for $\varepsilon> \varepsilon_c$, which for
$\varepsilon \rightarrow \varepsilon_c$ tends asymptotically to $\mu=0$
and $l^{1/2}\simeq 0.529$. For $\varepsilon \leq \varepsilon_c$ we have checked
numerically that the system (\ref{set}) has no solution.}
\label{2}
\end{figure}

In order to solve the system of eqs.(\ref{sdm}) and (\ref{sdlambda}) we have to distinguish two cases: a) $m_f=0$, and b) $m_f\neq0$.

a) If $m_f=0$ eq.(\ref{sdm}) is satisfied automatically, then we have checked numerically that eq.(\ref{sdlambda}) 
can be solved with respect to $\lambda$, for all the range of the free parameters $g$ and $m_b$. However, this class 
of solutions with $m_f=0,\lambda\neq 0$ is not accepted because of the IR divergences which arise when we go to second 
order approximation in momentum, as we explain in Appendix C. For this reason this class of solutions is not presented here.

b) If we assume that $m_f\neq0$ we can divide eq.(\ref{sdm}) by the factor $m_f^3$ to obtain eq.(\ref{mfgapeq}) in
the previous section. Then we can rescale the other parameters of
the theory with the renormalized mass of the scalar field $m_b$, to obtain the following dimensionless parameters:
\be \label{rescale}
\mu=\frac{m_f}{m_b}~~~~~~l=\frac{\lambda}{m_b^2}~~~~~~
\varepsilon=\frac{g}{2\pi m_b^3},
\ee
and the set of coupled equations to solve is, from eqs.(\ref{mfgapeq},\ref{sdlambda}),
\bea\label{set}
1&=&\varepsilon^2\int_0^\infty\frac{x^2~dx}{\tilde A_b\tilde A_f(\tilde A_b+\tilde A_f)} \nn
1&=&\frac{\varepsilon^2}{l}\int_0^\infty dx~x^8(x^2+l)\frac{2\tilde A_b+\tilde A_f}{\tilde A_b^3\tilde A_f(\tilde A_b+\tilde A_f)^2},
\eea
where
\be \tilde A_b=\sqrt{1+x^6}~~~~~~~~~~\tilde
A_f=\sqrt{\mu^6+x^2(x^2+l)^2}.
\ee
We solve the above algebraic system of equations numerically, and a unique solution for the pair $(l,\mu)$ is
obtained, if the dimensionless coupling $\varepsilon$ is larger than the threshold $\varepsilon_{c}\simeq 1.3263$.
The results for the parameters  $l^{1/2}$ and $\mu$ as a function of the dimensionless coupling $\varepsilon$
are presented in fig.(\ref{2}). We would like to stress that the singular point $\mu=0,\sqrt{ l}=0.529$ (for $\varepsilon=\varepsilon_{c}$) is not a solution of the system of equations (\ref{set}) because of the restriction $\mu\neq 0$. 
However the system of equations (\ref{set}) has solutions arbitrarily close to the singular point if $\varepsilon>\varepsilon_c$, 
while for $\varepsilon<\varepsilon_c$ we have no real solutions.

According to the above results Lorentz symmetry arises for the specific model we examine, in the IR limit when $p<<\lambda^{1/2}$.
Indeed, from the dispersion relation for the free fermion
\be
\frac{\omega^2}{\lambda^2}=\left(\frac{p^2}{\lambda}+1\right)^2 p^2+\frac{m_f^6}{\lambda^2},
\ee
and for $p<<\lambda^{1/2}$, we obtain
\be E^2\simeq
p^2+\tilde{m}_{f}^2,
\ee
where we define the rescaled parameters $E=\omega/\lambda$ and $\tilde{m}_{f}=m_f^3/\lambda$
that correspond to the fermion energy and mass with the correct dimensions $[E]=[\tilde{m}_f]=1$.

However, if $m_f\simeq\lambda^{1/2}$, the limit $p<<\lambda^{1/2}$ implies that $p<<m_f$,
and the behavior of the particle is then nonrelativistic, the kinetic energy of the fermion
is given by $p^2/2m_f$ (note that $m_f=\tilde{m}_f$ for $\lambda^{1/2}\simeq m_f$).
We observe in fig.(\ref{2}) that there is a small region for which we obtain a relativistic fermion,
in particular for $\varepsilon>\varepsilon_c$  when $\varepsilon$ is close to the critical value
$\varepsilon_c$, $l^{1/2}$ becomes significantly larger than $\mu$ .
For $\varepsilon>>\varepsilon_c$, the mass of the fermion increases and becomes comparable to $\lambda^{1/2}$,
this means that the relativistic behavior for fermions is restricted to a narrow set of values for the coupling.

The non-perturbative analysis with SD equations, in this article, was based on the following two
approximations: 1) ladder approximation, 2) first order approximation in momentum.

1) In the case of ladder approximation we make the assumption that the bare coupling $g$ is almost equal to the dressed coupling $\Theta$.
It seems that this consideration is consistent with our results, hence we do not
expect significant corrections in fig.(\ref{2}). In particular, in Appendix B we compute the one loop vertex diagram and show
that the dressed coupling  $\Theta^{(1)}$ receives small corrections and is therefore
close to the bare coupling $g$.

2) For $\varepsilon\simeq\varepsilon_c$, our approximation of taking into account only the first order in momentum in the propagators might not 
be reliable: higher order in momentum terms become significantly strong due to 
IR divergences for $m_f=0$,
hence different Ans\"{a}tze for the propagators should be considered, with a larger number of unknown parameters. This
would generate a considerably more involved numerical problem to solve and is beyond the scope of this article. However, as we explain in 
Appendix C, we believe that the effects of higher orders would smoothen the singular behaviour of the solution
when $\varepsilon\to\varepsilon_c$, and that the present singularity is 
rather an artifact arising from the first order approximation
in momentum.

Finally we would like to note that the issue of Lorentz symmetry restoration becomes problematic when we consider more
realistic models with different species of fermions. 
In particular, the authors of Ref.\cite{Iengo:2009ix} find that, although renormalization group flows 
reduce the difference in the velocities of different species towards the IR, this is not enough to 
achieve the experimental precision of the speed of light, unless we assume an unnatural fine-tuning between the 
Lorentz violating terms of the UV action of the model. These results are demonstrated in the framework of perturbation 
theory by considering several simple Lifshitz-type models, and we emphasize that the non-perturbative mechanism 
we propose in this paper suffers from similar problems, as we discuss further in the main part of this paper.
For example if we had considered an extension of our model with two species of fermions, with different Yukawa couplings $g_1$ and $g_2$, we would have encountered the generation of two first order kinetic terms with different $\lambda_1$ and $\lambda_2$ in the IR. 
This would mean that the limiting velocities of fermions would be significantly different in the IR, unless we had chosen $g_1$ extremely close to $g_2$. Such fine-tuning is not consistent with the logic of standard model where the Yukawa couplings are significantly different for different species of fermions.

\section{Conclusions}

We considered a 3+1 dimensional model with a Lifshitz-type fixed point ($z=3$) in which a scalar and a fermion field interact
via a Yukawa term. The effect of
dynamical mass generation, as well as the restoration of Lorentz symmetry in the IR limit was examined in the
framework of Schwinger-Dyson equations.

An interesting point in this model is that the interaction is super renormalizable,
and the only UV divergence present comes from the
scalar self-energy diagram. Note that, in contrast to the standard case ($z=1$)
in which the divergence is quadratic, the divergence in our model is
logarithmic, due to the higher powers of momentum in the propagators.
In order to absorb the UV divergence in our model, we introduce a bare mass for the
scalar such that our effective theory does not depend on the cutoff of the theory. 
On the other hand,
a bare fermion mass is not necessary in the action since a dynamical mass is generated quantum mechanically and is finite.

Note that the absence of quadratic divergences in the scalar self energy diagram sets the
hierarchy problem on a new basis,  as the scalar field mass flows logarithmically with the UV cutoff, see eq.(\ref{mb}). However,
the absence of quadratic divergences in our model can not be considered as a resolution to the 
hierarchy problem 
as we do not consider all the degrees of freedom of the Standard Model (see also \cite{Chao:2009dw}).

The ansatz for the scalar and fermion self energies is based on a linear approximation in the external momentum $k$,
and we do not discuss here the possibility of generating a Lorentz-invariant kinetic term for the scalar
field as this term would be of order
$k^2$. The equations arising from the Schwinger-Dyson approach are solved numerically and the
results are presented in fig.(\ref{2}). We find that there is a
critical value $g_c$ for the Yukawa coupling, above which Lorentz symmetry is restored and a mass is
generated in the low energy limit of the fermionic sector.

Note, that there are other marginal and relevant operators which can be included in the UV action, for example the 
interaction terms $g_1 \phi^2 \bar{\Psi}\Psi$, $g_2 (\bar{\Psi}\Psi)^4$, $g_3 \phi^4$, etc . However, had we 
included these terms in the bare action, the corresponding Schwinger-Dyson system of equations 
would have become exceedingly complicated  
and would in any case have lead us beyond the scope of this article. 
As a first step in a non-perturbative approach, we restricted our analysis to the most economical 
model, namely a Yukawa interaction, which is enough to demonstrate the dynamical generation of a 
mass and a first order kinetic term for the fermion. 

Finally, this non-perturbative mechanism for the restoration of Lorentz symmetry in models defined at a
Lifshitz point may be useful for the study of other
theories with immediate phenomenological interest, such as QED or Higgs models, which are proposed
for future investigation.

\vspace{1cm}

\nin{\bf Acknowledements} K. Farakos would like to thank D. Anselmi for useful discussions.
This work is partly supported by the Royal Society, UK, and partly by the National
Technical University of Athens through the Basic Research Support Programme 2008.

\section*{Appendix A: Schwinger-Dyson equation}

The partition function of the theory corresponding to the bare action (\ref{model}) is
\bea Z[j,\ol\eta,\eta]&=&\int{\cal
D}[\phi,\ol\psi,\psi]\exp\left\lbrace iS +i\int dtd^Dx\left( j\phi+\ol\eta\psi+\ol\psi\eta\right) \right\rbrace\nn &=&\exp\{iW[j,\ol\eta,\eta]\},
\eea
where $j,\ol\eta,\eta$ are the sources for $\phi,\psi,\ol\psi$ respectively, and $W$ is the connected graphs generator functional. 
The functional derivatives of the latter define the classical fields $\phi_c,\psi_c,\ol\psi_c$
\bea \frac{\delta W}{\delta j}&=&\frac{1}{Z}<\phi>\equiv\phi_c\nn
\frac{\delta W}{\delta\ol\eta}&=&\frac{1}{Z}<\psi>\equiv\psi_c\nn \frac{\delta W}{\delta\eta}&=&
-\frac{1}{Z}<\psi>\equiv -\ol\psi_c,
\eea
where
\be
<\cdots>=\int{\cal D}[\phi,\ol\psi,\psi](\cdots)\exp\left\lbrace iS +i\int dtd^Dx\left( j\phi+\ol\eta\psi+\ol\psi\eta\right) \right\rbrace.
\ee
The
proper graphs generator functional $\Gamma[\phi_c,\psi_c,\ol\psi_c]$ is defined as the Legendre transform of $W$,
\be
\Gamma=W-\int dtd^Dx\left(
j\phi_c+\ol\eta\psi_c+\ol\psi_c\eta\right),
\ee
where the sources have to be understood as functionals of the classical fields. It is easy to check
that
\bea\label{Gammaderivatives} \frac{\delta\Gamma}{\delta\phi_c}&=&-j\nn
\frac{\delta\Gamma}{\delta\psi_c}&=&\ol\eta\nn
\frac{\delta\Gamma}{\delta\ol\psi_c}&=&-\eta\nn
\frac{\delta^2\Gamma}{\delta\psi_c\delta\ol\psi_c}&=&-\left( \frac{\delta^2
W}{\delta\eta\delta\ol\eta}\right)^{-1}.
\eea
The first step for the derivation of a self consistent equation involving the dressed propagators and
vertex is to note that the functional integral of a functional derivative vanishes, such that
\be
\left<\frac{\delta S}{\delta\ol\psi}+\eta\right>=0.
\ee
Using the different derivatives (\ref{Gammaderivatives}), we obtain then
\be\label{eq1}
\frac{\delta\Gamma}{\delta\bar\psi_c}=\left(i\gamma^0\partial_t+(-\Delta)^\frac{z-1}{2} 
\left(i\gamma^k\partial_k\right)\right) \psi_c-\frac{g}{Z}<\phi\psi>.
\ee
The vertex, the bare and dressed fermion propagators are respectively
\bea
\Theta&=&\left(\frac{\delta^3\Gamma}{\delta\phi_c\delta\psi_c\delta\ol\psi_c}\right)_0\nn G_f^{-1}&=&
\left(\frac{\delta^2S}{\delta\psi\delta\ol\psi}\right)_0\nn {\cal G}_f^{-1}&=&\left(\frac{\delta^2\Gamma}{\delta\psi_c\delta\ol\psi_c}\right)_0,
\eea
where
the index 0 refers to vanishing fields, such that a functional derivative of eq.(\ref{eq1})
gives for the fermion self energy
\be\label{eq2}
\Sigma_f={\cal G}_f^{-1}-G_f^{-1}=-\frac{g}{Z}\left(\frac{\delta}{\delta\psi_c}<\phi\psi>\right)_0.
\ee
We then express $<\phi\psi>$ in terms of
derivatives of $W$:
\be
\frac{\delta^2W}{\delta j\delta\ol\eta}=-i\phi_c\psi_c+\frac{i}{Z}<\phi\psi>,
\ee
such that
\bea
&&\left(\frac{\delta}{\delta\psi_c}<\phi\psi>\right)_0\\
&=&-i\left(\frac{\delta}{\delta\psi_c}\frac{\delta^2W}{\delta j\delta\ol\eta}\right)_0\nn
&=&-i\left(\frac{\delta^3W}{\delta\eta\delta j\delta\ol\eta}~\frac{\delta\eta}{\delta j}\right)_0\nn &=&i\left(\frac{\delta}{\delta j}\left(\frac{\delta^2\Gamma}{\delta\psi_c\delta\ol\psi_c}\right)^{-1} \frac{\delta\eta}{\delta j}\right)_0\nn
&=&i\left(\left(\frac{\delta^2\Gamma}{\delta\psi_c\delta\ol\psi_c}\right)^{-1}
\left(\frac{\delta^3\Gamma}{\delta\phi_c\delta\psi_c\delta\ol\psi_c}\right) ~\frac{\delta\phi_c}{\delta j}~
\left(\frac{\delta^2\Gamma}{\delta\psi_c\delta\ol\psi_c}\right)^{-1} \frac{\delta^2\Gamma}{\delta\psi_c\delta\ol\psi_c}\right)_0\nn
&=&i{\cal G}_f\Theta\left(\frac{\delta j}{\delta\phi_c}\right)_0^{-1}
=-i{\cal G}_f\Theta{\cal G}_b,
\eea
and the Schwinger-Dyson equation for the fermion
self energy is finally, from eq.(\ref{eq2}),
\be
\Sigma_f=ig{\cal G}_f\Theta{\cal G}_b.
\ee
We stress that this equation is exact, and represents a resummation of all quantum corrections, since it involves the dressed
quantities on both sides of the equation.

\section*{Appendix B: Validity of the ladder approximation}

We check here that the ladder approximation is consistent and calculate corrections to the coupling constant.
This calculation is one-loop like, but it takes into account the fermion mass generated dynamically: this is a
similar approach to Schwinger Dyson equations, and avoids IR divergences.\\
This correction is given by the following three-point graph, for vanishing incoming momentum:
$$
\Theta^{(1)}=g+ig^3\mbox{tr}\int\frac{d\omega}{2\pi}\frac{d^3{\bf p}}{(2\pi)^3}
\left(\frac{\omega\gamma^0-{\bf p}^2 ({\bf p}\cdot\gamma)+m_f^3}{\omega^2-p^6-m_f^6+i\varepsilon}\right)^2
\frac{1}{\omega^6-p^6-m_b^6+i\varepsilon},
$$
and, after a Wick rotation,
$$
\Theta^{(1)}=g+\frac{g^3}{\pi^3}\int d\omega p^2dp\frac{-\omega^2-p^6+m_f^6}{(\omega^2+p^6+m_b^6)(\omega^2+p^6+m_f^6)^2}.
$$
In the phase where the fermion dynamical mass is generated, $m_f\simeq m_b$ (near the singular point of fig. \ref{2} but not very close to it), such that
\bea
\Theta^{(1)}&\simeq&g+\frac{g^3}{\pi^3}\int d\omega ~p^2dp\frac{-\omega^2-p^6+m_b^6}{(\omega^2+p^6+m_b^6)^3}\nn
&=&g+\frac{g^3}{4\pi^2}\int p^2dp\left(\frac{3m_b^6}{(p^6+m_b^6)^{5/2}}-\frac{2}{(p^6+m_b^6)^{3/2}}\right)\nn
&=&g+\frac{g^3}{12\pi^2m_b^6}\int_0^\infty dx\left( \frac{3}{(1+x^2)^{5/2}}-\frac{2}{(1+x^2)^{3/2}}\right)\nonumber
\eea
The last integral vanishes, and $\Theta^{(1)}\simeq g$. As a consequence, corrections to the coupling constant can be neglected,
and the ladder approximation that we used is justified.

\section*{Appendix C: Beyond first order approximation in momentum}

 In order to go beyond the first order approximation in momentum, we have to
expand the self energy diagrams up to second order in external momentum,
$$\Sigma_f({\bf k})=-m_f^3-\lambda_{f}({\bf k}\cdot\gamma)+Z_{f} k^2$$
$$\Sigma_b({\bf k})=m_b^6+\lambda_{b} k^2-m_{0}^{6}$$
In this way we introduce two more parameters $Z_{f}$ and $\lambda_{b}$,
and we have to solve numerically a system of four coupled equations with four
unknown parameters:
$$
\mu=m_f/m_b~~~~~~l_{f}=\lambda_{f}/m_{b}^2~~~~~~l_{b}=\lambda_{b}/m_{b}^{4}~~~~~~z_{f}=Z_{f}/m_{b},
$$
which would satisfy equations of the form
$$F_1(\mu,l_{f},l_{b},z_{f},g/m_{b}^3)=\mu$$
$$F_2(\mu,l_{f},l_{b},z_{f},g/m_{b}^3)=l_{f}$$
$$F_3(\mu,l_{f},l_{b},z_{f},g/m_{b}^3)=l_{b}$$
$$F_4(\mu,l_{f},l_{b},z_{f},g/m_{b}^3)=z_{f}$$
where the functions $F_1,F_2,F_3,F_4$ are integrals over the energy $\omega$ and the momentum $p$. Note that these functions
could be obtained by expanding the fermion and scalar self energies 
up to second order in momentum $k$, but the corresponding set of equations would lead to a much more involved numerical problem, and 
would not change the essential point, i.e. the dynamical generation of Lorentz-invariant terms.

Nevertheless, one can predict the effect these corrections would have, in the vicinity of the critical point of fig.\ref{2}. 
In our first order approximation in momentum, we disregarded the solutions $(\mu= 0,l_f\neq 0)$, whereas they
satisfy our set of self-consistent equations, since $l_f$ acts as an IR regulator when $m_f=0$. The reason we disregard
these solutions is the following: we checked that at  $m_f\to0$, the derivatives $d^2\Sigma_f({\bf k})/dk^2 $ and 
$d^2\Sigma_b({\bf k})/dk^2 $ become infinite whereas they should
be finite, since they are related to wave function renormalizations. As a consequence, we believe that 
the existence of this singularity is an artifact of the first order approximation in momentum and we expect that second 
order corrections would smoothen the singular behaviour of the dynamical mass.

\end{document}